# The role of doping and dimensionality in the superconductivity of $Na_xCoO_2$

M. Bañobre-López[1], F. Rivadulla[1*], R. Caudillo[3], M. A. López-Quintela[1], J. Rivas[2], J. B. Goodenough[3].

[1]*Physical Chemistry* and [2]*Applied Physics Departments, University of Santiago de Compostela, 15782-Santiago de Compostela (Spain)*
[3]*Texas Materials Institute, The University of Texas at Austin, Austin, TX 78712 (USA).*

We report a complete analysis of the formal $Co^{3+/4+}$ oxidation state in $Na_xCoO_2$, in the interval $0.31 \leq x \leq 0.67$. Iodometric titration and thermoelectric power confirm that a direct relationship between the Na content and the amount of $Co^{3+}$ cannot be established in this system. Creation of a significant amount of oxygen vacancies accompanies Na-ion deintercalation, keeping the formal Co valence at $3.45+$ for $x \leq 0.45$. To the light of new thermoelectric power data which reveals important differences between the hydrated (superconducting) and non-hydrated (non-superconducting) samples, we propose here that water plays an important "chemical" role beyond that of a spacer between the $CoO_2$ layers.

## INTRODUCTION

Superconductivity remains one of the most attractive and challenging areas of research in materials science. In this line, the recent report of superconductivity by Takada *et al.*[1] in hydrated $Na_xCoO_2$, has generated a huge activity and given an unprecedented impulse to the investigation of the fundamental properties of this system. However, an explanation of the role of the water in the occurrence of superconductivity in this material has remained elusive. Superconductivity was found after Na-ion deintercalation below x ≈ 0.42 and subsequent hydration.[1,2] $H_2O$ intercalates between the $CoO_2$ layers, increasing dramatically the lattice spacing and reducing the electronic dimensionality of the structure (the c-axis increases more than 50% over its original value). The proximity to a non-metallic phase establishes a possible comparison to the hole-doped cuprates, but the $CoO_2$ planes in $Na_xCoO_2$ adopt a 2D hexagonal symmetry to give the first example of a superconductive system with such geometry. On the other hand, the dome-shaped $T_C$ vs x curve reported by Schaak *et al.*[2] is now being challenged in a series of works[3,4] in which dimensionality is argued to have a strong influence on the appearance of superconductivity.

A systematic and extensive study of the formal valence state of Co versus Na and $H_2O$ contents is lacking. We report here that the Co valence is much lower than expected from the Na content. Both thermoelectric power and iodometric titration indicate that oxygen is progressively removed from the structure accompanying Na deintercalation. Our data point to an active role of $H_2O$ in the determination of the number of carriers in the $CoO_2$ layers, and the removal of the oxygen vacancies that strongly perturb the periodic potential.

## EXPERIMENT

Polycrystalline $Na_xCoO_2$ (x close to 0.67, determined experimentally) was prepared by a conventional solid state reaction. Dried $Na_2CO_3$ and $Co_3O_4$, were thoroughly mixed in a molar ratio Na / Co = 0.7 in an $Al_2O_3$ crucible that was placed directly in a preheated furnace at 850ºC to avoid Na evaporation and fired for 12h in air. A second heat treatment at 900ºC was carried out for 12 hours in air. After each heat treatment the sample was slowly cooled down to room temperature and reground. Na was chemically deintercalated from $Na_{0.67}CoO_2$ by stirring the powder in a $Br_2/CH_3CN$ oxidizing solution for five days at room temperature. Different $Br_2$ excesses (×1-×50) with respect to the stoichiometric amount needed to remove all of the Na were used in order to get a wide range of compositions. The products were washed several times with $CH_3CN$ and acetone, and then dried under vacuum. The Na and Co contents of the phases were determined by inductively coupled plasma optical emission spectroscopy (ICP-OES). The analysis confirms that the amount of Na decreases systematically as the excess of $Br_2$ increases. (Table 1).

**Table I.** Results of the chemical analysis of $Na_xCoO_2$. The two first samples correspond to the parent phase synthesized in slightly different conditions of temperature.

| Na content, x | %$Co^{3+}$ | Oxygen content, 2-$\delta$ | $Br_2$ excess |
|---|---|---|---|
| 0.690(1) | 89.0(1) | 1.90(5) | ------- |
| 0.673(7) | 77.3(7) | 1.95(2) | ------- |
| 0.450(4) | 61.0(4) | 1.92(2) | ×1 |
| 0.429(6) | 58.9(6) | 1.92(2) | ×5 |
| 0.402(5) | 56.6(5) | 1.91(1) | ×10 |
| 0.381(6) | 58.1(6) | 1.90(3) | ×20 |
| 0.369(5) | 53.8(5) | 1.92(3) | ×30 |
| 0.362(4) | 54.2(4) | 1.91(4) | ×40 |
| 0.322(4) | 52.2(4) | 1.90(3) | ×50 |
| 0.310(3) | 57.0(3) | 1.87(5) | ×100 |

The oxidation state of cobalt was determined for all the samples with the iodometric titration method. All the solutions were previously bubbled with Ar and the volumetric titration was carried out quickly to avoid any ambient oxidation of $I^-$. Superconducting samples were obtained by stirring $Na_xCoO_2$, x < 0.45, in water for two days at room temperature. Thermoelectric power was measured from 85 to 450 K in a home-made setup. Lattice parameters were calculated from the x-ray patterns after indexing all the reflections with Rietica.[5]

## RESULTS AND DISCUSSION

$Na_xCoO_2$ is a highly hygroscopic material, which makes it very unstable under ambient conditions. This and the unavoidable Na evaporation during the high-temperature synthesis, reduces control over the final stoichiometry and prevents a good characterization of the intrinsic and structural properties for a wide range of x. So, although the unit cell for the most stable phase, $x \approx 0.67$, is considered hexagonal with space group $P6_3/mmc$ (s. g. nº 194), hexagonal R3m (s. g. nº 160) and monoclinic C2/m (s. g. nº 12) have been reported for nearly the same composition.[6] Moreover, ordering of Na vacancies may change the symmetry. In the case of superconducting samples complexity and controversy is even bigger, because a mixture of fully and partially hydrated phases normally coexist in the same specimen[3].

We confirmed that the best fits of our powder diffractograms of the parent phase, $x = 0.67$, were achieved on the basis of a single phase with space group $P6_3/mmc$. On the other hand, in order to get a good fitting of the diffractograms corresponding to the Na-deintercalated samples, two new minority phases of the hexagonal space groups, $P6_3/m$ (s. g. nº 176) and $P6/m$ (s. g. nº 175), must be considered in addition to the $P6_3/mmc$ majority phase. The ratio of the c/a lattice parameters as a function of the Na content are shown in Fig.1 for the majority phase. A significant increase of the c-axis lattice parameter is observed with decreasing Na content (see the displacement of the (002) reflection in the inset of Fig.1) while the a-axis parameter decreases slightly, but continuously over all the sodium content range. This appreciable expansion of the unit cell along the c-axis is due to the decrease in bonding between the $CoO_2$ layers as Na is removed. Below $x \approx 0.4$, the expansion of the c-axis parameter is less marked, leading to an approximately constant c/a ratio down to $x \approx 0.3$.

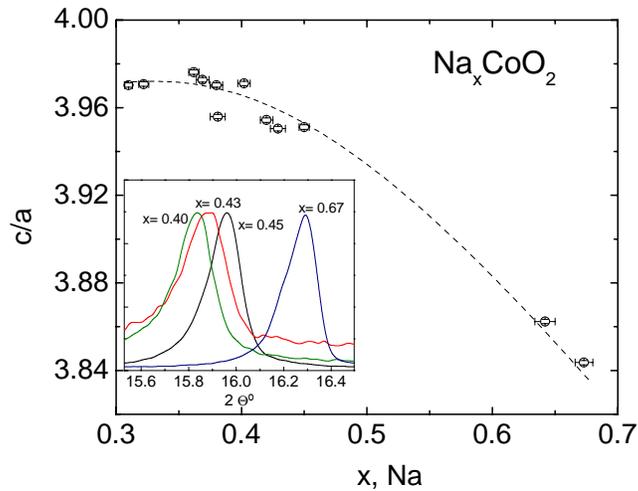

Fig, 1. Evolution of the c/a lattice parameters in $Na_xCoO_2$, for the majority phase $P6_3/mmc$. Inset: Displacement of the (002) reflection in $P6_3/mmc$ with Na content.

Several groups[3,4] recently reported that the superconducting critical temperature is not so dependent on the Na content as previously thought.[2] In fact, they show that $T_C$ is almost doping-independent in the optimal region for superconductivity. The superconducting state actually appears for a sufficiently large c/a ratio, *i.e.* when the 2D character is strong enough. However, a proper determination of the actual doping in terms of the true valence at the Co site is lacking. It is known that the related compound $Li_xCoO_2$ loses oxygen when $Li^+$-ion is removed[7], and some authors also demonstrated that the oxygen content is less than stoichiometric for some particular compositions of $Na_xCoO_2$.[8,9]

We have performed a careful chemical analysis of the oxidation state as a function of Na in a wide interval range, $0.310(3) \leq x \leq 0.69(1)$, including the optimal doping range for superconductivity. Reproducible iodometric titrations were consistent with an oxidation state of $Co^{3+/4+}$ that is considerably lower (a larger amount of $Co^{3+}$ and a smaller amount of $Co^{4+}$) than expected from the Na content determined by ICP-OES. The percentage of $Co^{3+}$ determined iodometrically is plotted in Fig. 2 versus the Na content determined by ICP-OES.

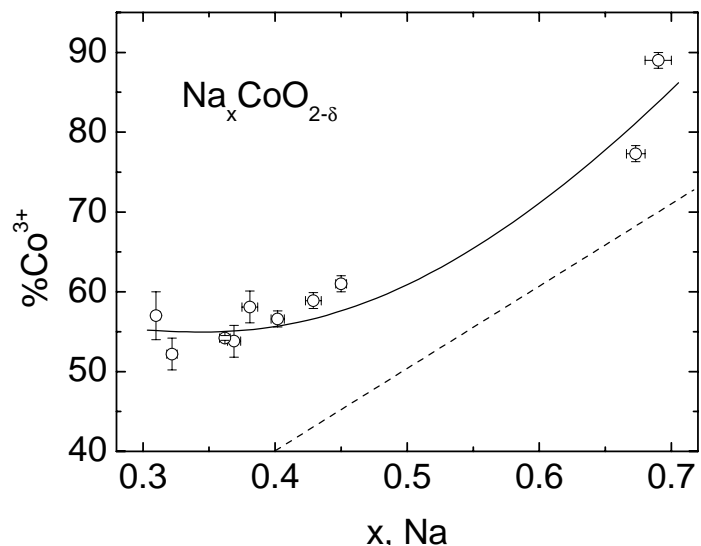

Fig. 2. The actual $Co^{3+}$ percentage as a function of the Na content. The solid line is a guide to the eye and the dotted line is the %$Co^{3+}$ expected on the basis of the Na content, supposing a perfect oxygen stoichiomety of 2.00.

The compound is always oxygen deficient, even at the largest doping probed in this work. That is consistent with the report by Sakurai et al.[10] of the presence of oxygen vacancies even for doping as large as x = 0.84. On the other hand, deintercalation of Na below x ≈ 0.4 has almost no effect in the final oxidation state of Co, which remains close to ≈ 3.45 +, down to at least 0.3, the lowest Na content probed in this work. The creation of an important number of oxygen vacancies below x ≈ 0.4 and the consequent reduction of the expected charge of the metallic ions is also reflected in an almost constant c/a ratio (see Fig. 1).

The loss of oxygen is made possible by a pinning of the $Co^{3+/4+}$: $t_{2g}$ band at the top of the $O^{2-}$:$2p^6$ band. In this case, a redox process between the pairs $Co^{3+/4+}$ and $O^{2-}/O_2$ can occur when holes are actually introduced into the oxygen band. Marianetti et al.[11] proposed an alternative mechanism in which hole-doping at the $t_{2g}$ band rehybridizates the $e_g$ and O:2p orbitals, which produces an effective hole transfer to the oxygen band and hence the same global redox process described above.

This redox process $Co^{3+/4+} \leftrightarrow O^{2-}/O_2$ makes it extremely difficult to reach a good control over doping of the $CoO_2$ layers through variation of the Na content.

Similar results of oxygen migration were reported in the related material $Li_xCoO_2$, where Venkatraman and Manthiram[7] demonstrated that the maximum oxidation state for Co is 3.5 + even after all the $Li^+$ was removed from the structure. The appearance of a large number of oxygen vacancies is also accompanied by a reduction in the c-axis parameter in the Li compound.

In order to confirm our results we have carried out systematic measurement of the thermoelectric power in the same samples analyzed chemically. Thermopower is the most sensitive electronic transport property of a metal: it is very susceptible to variations in the number of carriers and gives direct information of the asymmetry of the density of states around the Fermi energy. The results for the system $Na_xCoO_{2-\delta}$ are shown in Fig. 3.

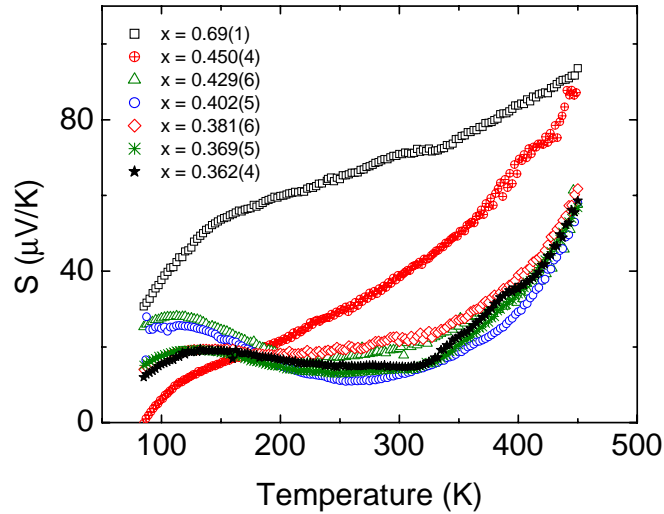

Fig. 3. Evolution of the temperature dependence of the thermopower with Na content. Below $x \approx 0.42$, the thermopower does not show a strong dependence on the Na/Co ratio due to the invariance of doping at the $CoO_2$ planes in this range of x.

The samples with high Na content present a thermoelectric power behavior similar to that previously reported by several authors.[12,13] The thermopower increases with temperature, but the value and the temperature dependence deviates from what in principle should be expected in a metal with such a low resistivity. As Na is removed, the thermopower decreases and becomes less temperature dependent, and in the low doping range it remains practically insensitive to variations in Na content. These measurements are fully consistent with the results obtained previously from iodometric titration analysis. The thermopower in the powder is dominated by the *ab*-plane[12] due to its higher conductivity; the influence of the dimensionality is low, so we are testing the variations in the number of charge carriers available for scattering in the electronically active $CoO_2$ planes (provided that the material remains metallic)[14]. Hence, the invariance of the thermopower in the low doping region corroborates the inefficiency of Na removal to introduce charge carriers in the $CoO_2$ layers due to the oxygen loss process.

Therefore, our results show unambiguously that a direct relationship between the Na content and the number of holes introduced into the $CoO_2$ planes does not exist in $Na_xCoO_2$. This effect is particularly dramatic at low doping ranges, where superconductivity is found. From this point of view, the variations in the $T_C$ found in samples with different values of x below 0.4,[2] cannot be related to a variation in the electronic charge in the $CoO_2$ planes. The parallelism with the cuprates in terms of out-of-plane doping control of the in-plane electronic charge is not completely valid in $Na_xCoO_2$. Oxygen plays an important role in the control of the doping of the $CoO_2$ planes in this system.

If doping remains almost unchanged across the so called "optimal doping range", the question to answer is: what is the origin of the variation of $T_C$ below x ≈ 0.4? Everything seemed to point towards a fundamental role of the structure by a modification of the dimensionality. However, in the original, non-superconductive compound, it is precisely below x ≈ 0.4 where the minimal changes are observed in the c/a ratio. This opens two possibilities for the role of $H_2O$ in the superconductivity:

1.- *Structural role*: $H_2O$ is a passive lattice spacer. Hydrated samples with lower Na content accept more $H_2O$ and become more 2D than those with higher Na content, favoring superconductivity. However, Schaak *et al.*[2] found that although the c-axis expands in the superconducting samples when lowering the Na content, the amount of intercalated $H_2O$ remains constant, about 1.3 molecule per formula unit, independent of x. Moreover, Shi *et al.*[15] reported a decreasing $T_C$ with an increasing c-axis lattice parameter, reflecting the strong controversy among the results.

2.- *Chemical role*: $H_2O$ plays an active role on the doping of the $CoO_2$ planes. If the oxygen of the intercalated $H_2O$ enters the oxygen vacancies of the $CoO_2$ planes (as occurs for bound $H_2O$ on the surface of an oxide particle), it gives its proton to the free interstitial water to create $(H_3O)^+$-ions. The $(H_3O)^+$ ions reduce the $CoO_2$ sheets like the $Na^+$ ions and the bound $O^{2-}$ ions decrease the strong perturbations of the periodic potential created by the vacancies. A reduction of the oxygen vacancies and an increase of the oxidation state of Co would explain the continuous increase in the c-axis lattice parameter that is observed in the hydrated samples as x is reduced, in spite of a constant amount of water.[2]

In order to check this active role of intercalated $H_2O$ in controlling the oxidation state of the $CoO_2$ planes, we have monitored the evolution of the thermopower as a non-superconductive sample is hydrated to render the superconductive phase. The results are plotted in Fig. 4. An increase of the thermopower in the superconducting sample is consistent with a substantial change in the oxidation of the $CoO_2$ planes due to $H_2O$ intercalation. Alternatively, the anisotropic expansion of the structure introduces the possibility of a change in the density of states at the Fermi Energy, $N(E_F)$, due to the overlapping of the $e^T$ and $a_1^T$ bands at $E_F$,[8,16] which could also contribute to the thermopower. To distinguish between these two possibilities, we have synthesized $K_xCoO_2$ (with identical structure to $Na_xCoO_2$ except for the interlayer space that is larger in the K sample) and compared the thermopower for two samples with almost identical composition (inset of Fig.4). The thermopower presents identical behavior in both materials irrespective of whether the interlayer spacer is $Na^+$ or $K^+$.

It is true that the increase in the c-axis parameter in the K samples with respect to the Na phase is much lower than in the hydrated superconductors, but the thermopower remains practically insensitive to the lattice spacer. This result makes us believe that a change in the oxidation state of the $CoO_2$ planes and the elimination of oxygen vacancies that strongly perturb the periodic potential by the introduction of $H_2O$ is realistic explanation of the role of the water in stabilizing the superconducting state.

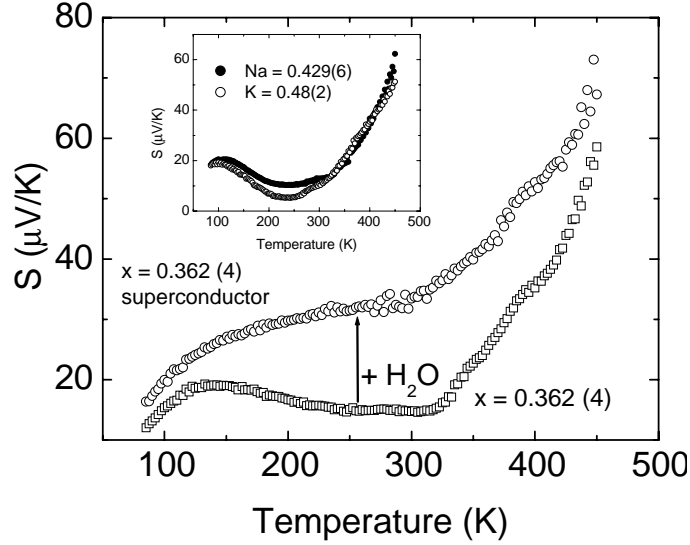

Fig. 4. Variation in the thermopower due to water intercalation. The $Na_{0.36}CoO_{2-\delta}$ was stirred in water for two days, after what was found to be superconductor below 3.7 K. Inset: Temperature dependence of the thermopower of $Na_{0.429(6)}CoO_{2-\delta}$ and $K_{0.48(2)}CoO_{2-\delta}$.

From our results it seems that the key for superconductivity is in a delicate balance between doping and dimensionality. Small variations in the doping and/or interlayer spacing will surely modify dramatically the $N(E_F)$ through the narrowing and relative position of the antibonding $a^T$ band with respect to the $E_F$. The narrowing of this band will surely increase the Pauli susceptibility in the hydrated samples with respect to the non hydrated ones, and in a simple BCS picture of the problem, will modify $N(E_F)$ which enters at the electron-phonon coupling parameter, and hence $T_C$.

CONCLUSIONS

We have demonstrated that below $x \approx 0.4$ in $Na_xCoO_2$, the formal oxidation state of Co remains constant around 3.45 +, which indicates oxygen loss from the $CoO_{2-\delta}$ layers. We have argued that the water insertion allows the introduction of oxygen into the oxygen vacancies and that the acidic character of a $Co^{4+}/Co^{3+}$ oxide would release hydrogen from $OH^-$ groups on the $CoO_{2-2\delta}(OH)_{2\delta}$ layers to the interstitial $H_2O$ to create $H_3O^+$ ions in the Na layers between $CoO_2$ layers. In this way the water removes the perturbation of the periodic potential in the $CoO_2$ layers and oxidizes them more deeply than is possible by $Na^+$-ion removal alone.

The role of lattice dimensionality is probably important in the occurrence of superconductivity in the hydrated samples by a modification of $N(E_F)$, but our experiments point to $H_2O$ molecules playing a role other than a passive lattice spacer, with a direct implication in the control of the effective doping at the $CoO_2$ planes.

**Acknowledgments**. This work was financed by FEDER project MAT2001-3749, MC&T, Spain. M. B-L. also thanks the MC&T for a FPU grant, and F. R. for support under program Ramón y Cajal. J. B. G. acknowledges the NSF and the Robert A. Welch Foundation of Houston, TX, for financial support.